\documentclass[apjl]{emulateapj}
\shorttitle{Local Pancake Defeats Axis of Evil}
\shortauthors{Vale}

\begin{document}

\title{Local Pancake Defeats Axis of Evil}

\author{Chris Vale}
\affil{Theoretical Astrophysics, Fermi National Accelerator Laboratory, Batavia, IL 60510 \\
Physics Department, University of California, Berkeley, CA 94720}
\email{cvale@fnal.gov}

\begin{abstract}
Among the biggest surprises revealed by COBE and confirmed by WMAP 
measurements of the temperature anisotropy of the CMB are the anomalous 
features in the 2-point angular correlation function on very large 
angular scales.  In particular, the $\ell = 2$ quadrupole and 
$\ell = 3$ octopole terms are surprisingly planar and aligned with one 
another, which is highly unlikely for a statistically isotropic Gaussian random field, and the axis 
of the combined low-$\ell$ signal is perpendicular to the ecliptic plane and the plane 
defined by the dipole direction.  Although this $< 0.1 \%$
3-axis alignment might be explained as a statistical fluke, it is certainly an uncomfortable 
one, which has prompted numerous exotic explanations as well as 
the now well known  ``Axis of Evil'' (AOE) nickname.  Here, we present a novel 
explanation for the AOE as the result of weak lensing of the CMB 
dipole by large scale structures in the local universe, and demonstrate that the 
effect is qualitatively correct and of a magnitude sufficient to fully explain the anomaly.  
\end{abstract}


\section{Introduction}

The full-sky maps obtained by
the Wilkinson Microwave Anisotropy Probe (WMAP) in its first year of
observation have revolutionized the study of the CMB sky \citep{Bennett_foregr,
Bennett2003, Hinshaw2003,Spergel2003}.  One of the many exciting uses of these 
maps is the window they offer us on the 
physics of the early universe \citep{Peiris2003} and the 
standard inflationary model.  This picture predicts a statistically isotropic 
Gaussian random CMB temperature anisotropy to an excellent approximation, so that 
WMAP has enabled considerable study of non-Gaussianity 
and statistical anisotropy of the CMB \citep[see e.g.][and references therein]{CopiHSS05}.  

Although WMAP data has largely been consistent with the standard cosmological picture, 
some challenges to the standard model have emerged.  
Among the most interesting occur on the largest angular scales, and are 
therefore called the  ``low-$\ell$ anomalies''.  The first of these, the surprisingly small 
quadrupole moment, was noted a decade ago in the COBE data 
\citep{DMR4_Ctheta} and has recently been confirmed by WMAP \citep{Spergel2003}.  
In addition, it is now known that the octopole is highly planar and 
aligned with the quadrupole \citep{SchwarzSHC04,OliveiraTZH04}, and that the planes defined by 
these multi-poles is perpendicular to both the dipole and the ecliptic.  

The original low quadrupole anomaly has long been dismissed as either 
the result of some residual systematic error or as a statistical fluke.  However, the higher 
quality of data now available from WMAP strongly challenge the residual systematic 
explanation \citep{TegmarkOH03}, and while a statistical fluke cannot be ruled out, 
the odds against are uncomfortably long.  In one recent study, \cite{CopiHSS05} have used the 
multi-pole vector formalism to show that a purely accidental alignment 
is unlikely in excess of $99.9 \%$.  They have also shown that most of the $\ell = 2$ 
and $\ell = 3$ multi-pole vectors of known Galactic foregrounds are located far away 
from those observed in WMAP data, strongly suggesting that residual 
contamination by foregrounds which are currently included in the analysis 
is not a viable explanation.  It is precisely this combination of a complete lack of any known 
systematic error, and long odds against random alignment that has earned the 
low-$\ell$ alignment anomaly the nickname  ``Axis of Evil'' \citep[AOE;][]{LandM05a}.  
The axis appears to possess mirror symmetric properties \citep{LandM05b} and may be 
related to the north-south asymmetry in the angular power spectrum \citep{Eriksen_asym}.

The fact that the AOE alignment includes sources of both cosmological and local origin 
suggests that the explanation might be found in some sort of 
interaction with local structure which is not already included in the foreground analysis.  
If this is the case, then the fact that the 
AOE points in the direction of the Virgo cluster \citep{OliveiraTZH04} is certainly 
intriguing, and has led some to suggest \citep{AbramoS03} that 
the SZ \citep{SunyaevZ72} imprint of the local supercluster might help 
explain at least some of the AOE.  Although this idea is initially attractive, the SZ 
is probably 3 or 4 orders of magnitude too small 
to do the trick \citep{DolagHRM05}.  

Local effects remain an attractive option, especially since the mass distribution is so highly anisotropic.  
The observed motions of galaxies in the local universe have revealed a bulk flow in the direction of the 
``Great Attractor'' region \citep{Lynden-Bell_88}.  The Local Supercluster, in which galaxies lie preferentially in the 
supergalactic plane \citep{de Vaucouleurs58}, dominates as we move closer to the Earth, and of course the 
Milky Way itself dominates on still smaller scales.  

These all contribute to the velocity of the Earth relative to the background, which gives rise to the CMB dipole term.  
The dipole is more than two orders of magnitude larger than the quadrupole and octopole, and is 
by far the largest of the temperature anisotropies; however, since the dipole 
arises from the peculiar motion of the observer with respect to the background, it is clearly of 
non-cosmological origin, and is therefore measured and subtracted from maps of the CMB.  Here, we 
show that this subtraction is imperfect.  Weak gravitational lensing by local large scale structures will 
coherently deform the initially perfect dipole, causing a leakage of power at the sub-percent level into 
other low-$\ell$ moments.  

We suggest, to our knowledge for the first time, that it is this lensing induced 
mixing of power from the CMB dipole that is the root cause of the AOE.
We will introduce the basic idea, along with some relevant background in 
Section \ref{sec:2}, where we will show that the explanation is natural and qualitatively 
correct.  In Section \ref{sec:3}, we will present the results of a toy model simulation 
which demonstrate the effect to be of the right order of magnitude, and we will conclude 
in Section \ref{sec:4} 
with some discussion of the cosmological implications of the reinterpreted large scale 
CMB sky in the event that the lensed dipole ultimately proves to be the right answer.

\section{Weak Lensing of the CMB Dipole} \label{sec:2}

Weak lensing of the CMB has long been a topic of interest to cosmologists 
\citep[e.g.][]{Seljak96, ZaldarriagaS98,Hu00,ChallinorL05}.  
The effect is both simple and inescapable; 
all light which reaches us from the surface of last scattering (or any other source, 
for that matter) is deflected from its original path by the weak gravitational lensing 
interaction with the matter distribution along the line of sight
\citep[see][for a comprehensive review]{BartelmannS01}, and no exceptions are 
made for photons from the CMB dipole.  Although the dipole owes its existence to the 
motion of the observer with respect to the background, this makes no difference from the 
perspective of someone in the same frame of reference as the observer; one side 
of the universe is simply hotter than the other, and this anisotropy will be lensed.  For the case of an observer comoving with the 
Earth, the observed dipole term will be that measured by WMAP 
\citep{Bennett2003}, which is more than two orders of magnitude larger than 
the quadrupole term (and is by far the largest anisotropy in the CMB), so that 
even sub percent level scatter will strongly effect the low-$\ell$ moments.  Also, because 
the dipole is coherent over the whole sky, it will couple best to lensing effects that 
are also coherent over much of the sky, so that local structures will be the dominant  lenses.  

This cannot be accounted for by simply subtracting the measured dipole; lensing will scatter 
the initially pristine dipole into something that is only \emph{almost} a perfect dipole, so that 
if we fit a dipole to the measured sky and subtract it, we are going to be stuck with 
a residual.  It is this residual which we believe is a likely culprit to explain the AOE, as we will 
now explain.  In the following, we will adopt a frame of reference that moves with the lens and 
observer, work in comoving coordinates $\chi$, set $c=1$, and assume a flat universe.  

\subsection{Lensing and the CMB}

The notion that gravitational lensing by large scale structure will scatter power 
between $\ell$-modes of the CMB is certainly not a new one \citep[]{Seljak96}, and while we 
will consider effects on much larger than usual angular scales, we will 
begin our discussion in the usual way.
While en route to us, the photons of the primary CMB are gravitationally lensed 
\citep[see e.g.][for a recent discussion of CMB lensing]{ChallinorL05} by the mass distribution 
along the line of sight all the way from the epoch of last scattering to the present, so that 
the photons we see have been displaced from their original position on the sky by an angle  
\begin{equation} \label{eq:alpha}
\vec\alpha (\hat{n}) = 2 \int_{0}^{\chi '} \!\! d \chi \, {\chi' - \chi \over \chi'}
\nabla_{\!\! \perp} \Phi
\end{equation}
where $\hat{n} = (\theta,\phi)$ is the angular position on the sky, 
$\nabla_{\!\! \perp}\Phi$ is the spatial gradient of the 
gravitational potential perpendicular to the path 
$d \chi$ of a given light ray, and $\chi '$ is the distance to the surface of 
last scattering.  The scattering can be well described as the 
angular gradient of an effective projected potential $\Psi$, so that 
if the source is much farther away than the lens, as is the case for us, the deflection is
\begin{equation} \label{eq:psi}
\vec \alpha = \nabla \left [ 2 \int _{0}^{\chi '} \!\! d \chi \, {\Phi \over \chi} \right ]
\end{equation}
where $\nabla$ is with respect to angular coordinates so that $\nabla = \chi \nabla_{\perp}$, and the quantity inside the 
brackets defines $\Psi$.  
The observed temperature $T$ is a simple re-mapping of the primordial value $T'$ 
\begin{equation} \label{eq:Remap}
T(\hat{n}) = T'(\hat{n} - \vec\alpha)
\end{equation} 
If we expand 
equation \ref{eq:Remap} to linear order, then the observed temperature is simply 
approximated as
\begin{equation} \label{eq:linear}
T(\hat{n}) = T'(\hat{n}) - \nabla T \cdot \nabla \Psi
\end{equation}
\begin{figure*} [!t]
{\includegraphics[width=\textwidth]{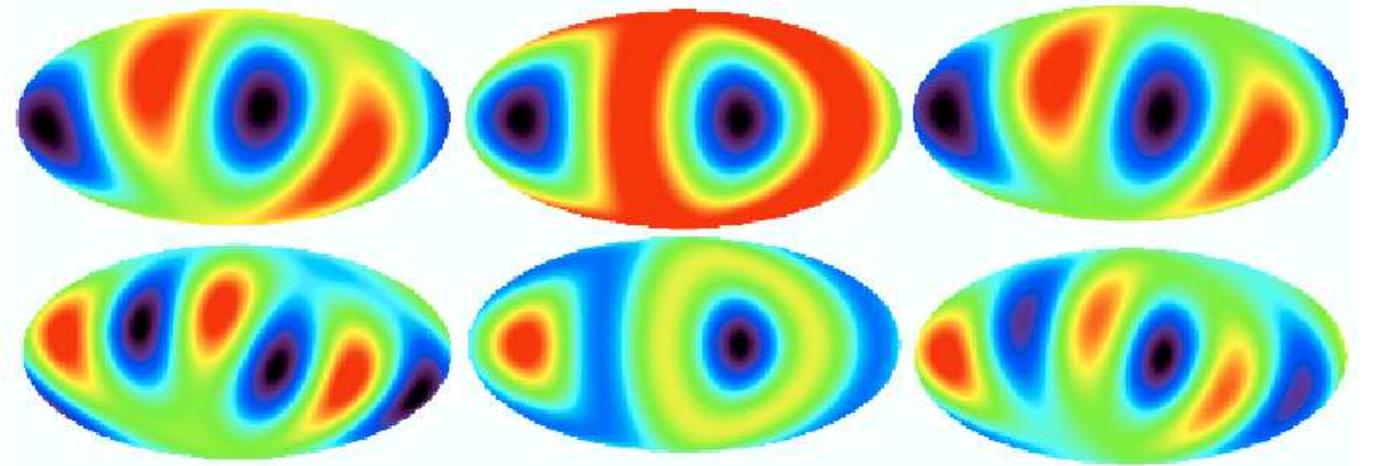}} \\
%{\includegraphics[width = 2.4in]{wquad} \includegraphics[width = 2.4in]{shapquad} \includegraphics[width = 2.4in]{quad_rigged} \\
%\includegraphics[width = 2.4in]{woct} \includegraphics[width = 2.4in]{shapoct} \includegraphics[width = 2.4in]{oct_rigged}}
%
\caption{The CMB temperature quadrupole (top panels) and octopole (bottom panels).  From left to right: the observed moments from 
WMAP \citep[as reported by][]{OliveiraTZH04}, the moments from the simplest toy model with one attractor, and the moments 
from the 6 attractor model.  Note that the 
color scale of the plots has been adjusted for display, spanning $\pm 35$mK for the WMAP octopole, $\pm 10 \mu$K 
for the one attractor octopole, and $\pm 15 \mu$K for the rest. }
\label{fig:fig1}
\end{figure*}

\noindent
where for our purposes $\nabla T$ is dipole as seen by a given lens.  
Equations \ref{eq:psi} and \ref{eq:linear} highlight some important points about CMB lensing.  
First, the lensing deflection falls off slowly, as $\alpha \sim 1 / \theta$, so that the deflection effect of large concentrations 
of mass extends much further on the sky than lensing induced shear and magnification \citep[e.g.][]{BartelmannS01}.  
Second, the magnitude of the deflection from a given overdensity is weighted as $\alpha \sim M / \chi$, so that the relatively 
local universe will contribute substantially to the overall deflection field; points one and two are crucial given our need for deflections 
which are coherent over huge fractions of the sky.  Third, since $\nabla T$ and $\nabla \Psi$ are proportional to the velocity and mass, 
respectively, the scatter will be maximized around large momentum overdensities.  
Finally, power which is scattered from the CMB dipole will be locally perpendicular 
to its gradient, establishing a preferred axis.  

\subsection{Lensing and the integrated Sachs-Wolfe effect}

We have so far considered the gravitational interaction of the CMB with lenses that are comoving with the observer, 
and we will now generalize this result.  We will start by considering the effect in the rest frame of the 
CMB, where it is known as the integrated Sachs-Wolfe (ISW) effect.  We note that 
a number of people have commented to the author that the ISW is in some ways a more intuitive description.  
While we do not share this view, it's certainly worth pointing out the relation, dictated by the equivalence principal, 
between the two (in our opinion) equally valid descriptions.

Let us consider a photon with a 4-momentum $p^{\mu} = E_0(1, \hat{n})$ traversing a potential which 
is changing in time, so that the energy a photon gains when falling into the well is not equal 
in magnitude to that lost when climbing out.  The observed potential 
will be described by a metric with $\Phi \! \left ( x - (V_{\rm lens} - V_{\rm obs}) t \right )$, so that 
(working to first order) the observed change in energy is
\begin{equation} \label{eq:isw}
dE_{\rm ISW} = 2 \, d \chi \, E_0 \, 
(\vec{V}_{\rm lens} -\vec{V}_{\rm obs}) \cdot \nabla_{\!\! \perp} \Phi_{\rm lens} 
\end{equation}
where $\vec{V}_{\rm obs}$ is the observer's velocity and 
$\nabla_{\perp}$ implies the direction perpendicular to the path of the photon.  To compute 
the dipole lensing term, we take the dot product of the change in 
4-momentum of the photon with the 4-velocity the observer, so that 
$dE_{\rm lens} = E_0 \vec{V}_{\rm obs} \cdot d\hat{n}$, and 
\begin{equation} \label{eq:diplens}
dE_{\rm lens} = 2 \, d \chi E_0 \, \vec{V}_{\rm obs} \cdot \nabla_{\! \perp} \Phi_{\rm lens}
\end{equation}
Adding equations \ref{eq:isw} and \ref{eq:diplens}, integrating, and noting that 
$E_0 V_{\rm lens}$ is the dipole magnitude seen by the lens, we arrive at 
\begin{equation} \label{eq:de}
\Delta E = 2 \, \nabla T_{\rm dipole} \cdot \vec{\alpha}
\end{equation}

\section{Results from a Toy Model} \label{sec:3}

We compute the lensing distortion of the CMB dipole in two steps.  First, for a given mass distribution, we 
compute the deflection field $\vec \alpha (\hat{n})$ using the 3-d ray-tracing method described in 
\cite{ValeW03} over the whole sky.  This method first computes the 3-d gradient of the gravitational potential 
on a 3-d grid, maps the result onto the points of a 3-d spherical grid, and selects the components which 
are locally perpendicular to the line of sight.  This gives us $\nabla_{\perp} \Phi (\chi,\hat{n})$ from 
equation \ref{eq:alpha}, so that $\vec \alpha$ is given by a sum over $\chi$ at each position on the 
spherical surface $\hat{n} = (\theta, \phi)$. 
We then generate the temperature map for the lensed dipole  on the 2-d spherical grid 
by following the prescription of 
equation \ref{eq:Remap}, in which we compute the dipole temperature at position 
$(\hat n - \vec \alpha)$ and apply it to position $\hat n$.  The spherical harmonic coefficients 
$a_{\ell m}$'s of the new field are then computed, and the $\ell > 1$ terms are observed to no longer be zero.

However, before we can employ this technique, we must first decide on a mass 
distribution from which to derive the lensing deflection.  This is a challenging proposition given the 
current level of uncertainty in the local distribution of 
dark matter.  The closest superclusters such as the Local Supercluster, the Centaurus Wall, 
the Perseus-Pisces chain, and the Great Attractor, all intersect the Milky Way ``Zone of Avoidance", 
so that Galactic extinction makes a full probe of their properties currently problematic 
\citep[see e.g.][for a recent discussion]{Kraan-Korteweg05}.  
Furthermore, even if this obstacle were somehow overcome, our ability to probe the dark matter 
distribution, using optical tracers or any other method, is highly uncertain.  We are 
at present unable to confidently state even the source of the mass dipole moment 
responsible for the bulk flow of galaxies in the local universe, with opinions divided between the so called 
Great Attractor region (centered on the ultra-massive Norma cluster ACO 3627), which is 
roughly $45 h^{-1}$Mpc from us, and the Shapley Supercluster, which is 
$100 h^{-1}$Mpc further.  

Because we are interested in a proof of concept analysis, we avoid the painstaking complexities involved 
in modeling the local universe, which cannot in any case be done with any more certainty than current observations 
allow.  Instead, we employ a toy model motivated by the flow of mass in the local universe.  
We begin by placing an overdensity near the Shapley concentration at a distance of $200 h^{-1}$Mpc, with a fwhm of 
$100 h^{-1}$Mpc and a 
peak density five times the mean density of the universe, which is consistent with the 
observed galaxy overdensity in the region \citep[][]{Drinkwater_04}.  The overdensity is sufficient to induce a 
peculiar velocity at the Milky Way of roughly $600$ km/s,  which increases to over $2000$ km/s near the peak of 
the mass; remarkably, this is also consistent with recent observations \citep{KocevskiE05}.  The magnitude of the induced 
quadrupole and octopole are similar to those found in the WMAP data, about equal for the former and 
one third for the latter, while the orientation of these (Figure \ref{fig:fig1}) is also remarkably 
close given the simple model employed. 

As a demonstration (Figure \ref{fig:fig1}), the author has attempted to recreate the observed $\ell = 2,3$ modes ``by hand'' by placing several 
overdensities and voids at a distance of $200 h^{-1}$Mpc, smoothed now on a $150 h^{-1}$ Mpc scale, with 
no net mass added.  The resulting 
density field peaks at three times the mean density in the direction of the Shapley concentration, where the velocity 
of the flow also peaks at just over $2000$ km/s, and the induced peculiar velocity at the Galaxy is $700$ km/s.  Although 
this toy is clearly not meant to be taken seriously, it is instructive that no particularly unrealistic mass 
concentrations are required to reproduced the observed modes.  It is also instructive to note that   
most of the signal comes from structures located far away from any overdensities; the velocity field effects a huge volume of 
space, which is typically at the mean density, and it is this extended region that contributes the lion's share of the effect.

We note that it is also relatively simple to create distributions which contribute power to alternating modes;  for example, 
a dumbell configuration, with the Galaxy at the center, contributes power to the $\ell = 2,4,6 \dots$ modes, and none at all 
to the odd modes.  We also note that because power is preferentially added in the direction of the 
flow, this may fully or partially explain the north-south power asymmetry \citep{Eriksen_asym}.  However, these effects depend strongly on 
the specific distribution of mass, so we have not attempted to quantify them here.

\section{Discussion} \label{sec:4}

We have introduced a novel explanation for the alignment of the axes defined by the planes of the 
CMB quadrupole and octopole, with each other and with the CMB dipole, which has become known as the 
``Axis of Evil''.  We find that the CMB dipole induced by the flow of local structures 
is scattered by gravitational interaction into the higher moments, and that although the effect on the dipole is at the sub-percent level, 
it is of order one for the higher moments, whose measured values are roughly two orders of 
magnitude smaller than the typical dipole seen by local structure.  We have shown that dipole 
lensing is a natural mechanism to align axes and create an asymmetric distribution of power, 
and have explicitly demonstrated that it is 
not out of bounds to attribute the entire measured CMB quadrupole to dipole lensing.  Consequently, it is 
highly unlikely that dipole lensing can be safely ignored in attempts to measure the large angle power in the CMB. 

While we \emph{may} have mitigated the alignment problem, the mechanism will on average 
add power to the low-$\ell$ moments, so that the 
low quadrupole anomaly is stronger than ever.  Low-$\ell$ modes are profoundly affected 
by the influence of Dark Energy (DE) through the integrated Sachs-Wolfe effect.  At late times, most models 
predict an epoch of equality between DE and Dark Matter (DM), after which DE causes an accelerated expansion.  
This expansion causes potential wells on scales not yet virialized to decay, causing a differential in the 
blue and red-shift photons experience as they traverse in and out of the wells.  Overestimating the low-$\ell$ power 
will lead to an overestimation of this differential, and lead to erroneous conclusions about the epoch of DE-DM equality, 
and therefor about dark energy parameters.  

Overestimation of the low-$\ell$ modes of the temperature may cause still other problems.  One of the many 
important observations in the WMAP data is that there is more than the expected amount of power 
in the low-$\ell$ modes of the cross correlation, $C_{\ell}^{TE}$, between E-polarized modes and temperature, 
which has been used to infer a higher than expected re-ionization depth \citep{Bennett2003}.  If much of the power 
in the observed data is due to the dipole and lensing, then it may be that the measured $C_{\ell}^{TE}$ is biased high, 
which would alter the predicted ionization depth.

Whatever their cause ultimately proves to be, the low-$\ell$ anomalies in the CMB temperature remain 
one of the fertile areas of research beyond the standard cosmological model, and just might prove to be the 
right window to look through to glimpse the physics driving the expansion of the universe.

\emph{CV acknowledges useful conversations with Martin White, Joanne Cohn, Wayne Hu, and owes a special debt 
to Alexia Schulz.  He also wishes to thank the many people 
who took the time to post useful comments on the cosmocoffee.info.  This research was supported by the NSF and NASA.}


\begin{thebibliography}{99} 

\bibitem[Abramo \& Sodre(2003)]{AbramoS03}
Abramo R., Sodre J., 2003, preprint [astr-ph/0312124]

\bibitem[Bartelmann \& Schneider(2001)]{BartelmannS01}
Bartelmann, M., \& Schneider, P., 2001, Phys. Rep., 340, 291-472

\bibitem[{{Bennett} {et~al.}(2003{\natexlab{a}})}]{Bennett_foregr}
{Bennett} C.~L., {et~al.}, 2003{\natexlab{a}}, \apjs, 148, 97

\bibitem[{{Bennett} {et~al.}(2003{\natexlab{b}})}]{Bennett2003}
{Bennett} C.~L., {et~al.}, 2003{\natexlab{b}}, \apjs, 148, 1

%\bibitem[{Cayon {et~al.}(2005)Cayon, Jin, \& Treaster}]{Cayon2005}
%Cayon L., Jin J., Treaster A., 2005, astro-ph/0507246

\bibitem[Challinor \& Lewis(2005)]{ChallinorL05}
Challinor A., Lewis A, Phys. Rev. D, 2005, 71, 103101

%\bibitem[{Chiang {et~al.}(2003)Chiang, Naselsky, Verkhodanov, \&
%  Way}]{Chiang2003}
%Chiang L.-Y., Naselsky P.~D., Verkhodanov O.~V., Way M.~J., 2003, \apj, 590,
 % L65

\bibitem[Copi et al.(2005)]{CopiHSS05}
Copi C., Huterer D., Schwarz D., Starkman G., preprint [astro-ph/0508047]

%\bibitem[{Cruz {et~al.}(2005)Cruz, Martinez-Gonzalez, Vielva, \&
 % Cayon}]{Cruz2004}
%Cruz M., Martinez-Gonzalez E., Vielva P., Cayon L., 2005, \mnras, 356, 29

\bibitem[de Oliveira-Costa et al.(2004)]{OliveiraTZH04}
de Oliveira-Costa A., Tegmark M., Zaldarriaga M., Hamilton A., 
Phys. Rev. D, 2004, 69, 063516

\bibitem[de Vaucouleurs(1958)]{de Vaucouleurs58}
de Vaucouleurs G., 1958, AJ, 63, 253

\bibitem[Dolag et al.(2005)]{DolagHRM05}
Dolag K., Hansen F., Roncarelli M., Moscardini L., 2005, MNRAS, 361, 753

\bibitem[Drinkwater et al.(2004)]{Drinkwater_04}
Drinkwater M., Parker Q., Proust D., Slezak E., Quintana H., 2004, PASA, 21, 89

%\bibitem[{Donoghue \& Donoghue(2005)}]{Donoghue}
%Donoghue E.~P., Donoghue J.~F., 2005, Phys. Rev., D71, 043002

%\bibitem[Eriksen et al.(2005)]{Eriksen_Npt}
%Eriksen H., 2005, \apj., 622, 58

\bibitem[{{Eriksen} {et~al.}(2004){Eriksen}, {Hansen}, {Banday},
  {G{\' o}rski}, \& {Lilje}}]{Eriksen_asym}
{Eriksen} H.~K., {Hansen} F.~K., {Banday} A.~J., {G{\' o}rski} K.~M., {Lilje}
  P.~B., 2004, \apj, 605, 14

%\bibitem[{Hajian \& Souradeep(2003)}]{Hajian2003}
%Hajian A., Souradeep T., 2003, \apj, 597, L5

%\bibitem[{Hajian \& Souradeep(2005)}]{Hajian2005}
%Hajian A., Souradeep T., 2005, astro-ph/0501001

%\bibitem[{{Hajian} {et~al.}(2005){Hajian}, {Souradeep}, \&
%  {Cornish}}]{Hajian2004}
%{Hajian} A., {Souradeep} T., {Cornish} N., 2005, \apj, 618, L63

%\bibitem[{{Hansen} {et~al.}(2004{\natexlab{a}}){Hansen}, {Banday}, \& {G{\'
 % o}rski}}]{Hansen_isotropy}
%{Hansen} F.~K., {Banday} A.~J., {G{\' o}rski} K.~M., 2004{\natexlab{a}},
%  \mnras, 354, 641

%\bibitem[{{Hansen} {et~al.}(2004{\natexlab{b}}){Hansen}, {Cabella},
 % {Marinucci}, \& {Vittorio}}]{Hansen_asym}
%{Hansen} F.~K., {Cabella} P., {Marinucci} D., {Vittorio} N.,
 % 2004{\natexlab{b}}, \apj, 607, L67

\bibitem[{{Hinshaw} {et~al.}(1996){Hinshaw}, {Branday}, {Bennett}, {G{\'
  o}rski}, {Kogut}, {Lineweaver}, {Smoot}, \& {Wright}}]{DMR4_Ctheta}
{Hinshaw} G., {Branday} A.~J., {Bennett} C.~L., {G{\' o}rski} K.~M., {Kogut}
  A., {Lineweaver} C.~H., {Smoot} G.~F., {Wright} E.~L., 1996, \apj, 464, L25

\bibitem[{{Hinshaw} {et~al.}(2003{\natexlab{a}})}]{Hinshaw2003}
{Hinshaw} G., {et~al.}, 2003{\natexlab{a}}, \apjs, 148, 135

\bibitem[Hu(2000)]{Hu00}
Hu W., Phys. Rev. D, 2000, 62, 043007

%\bibitem[{{Komatsu} {et~al.}(2003)}]{Komatsu2003}
%{Komatsu} E., {et~al.}, 2003, \apjs, 148, 119

\bibitem[Kocevski \& Ebeling(2005)]{KocevskiE05}
Kocevski D., Ebeling H., submitted to ApJ, preprint [astro-ph/0510106]

\bibitem[Kraan-Korteweg(2005)]{Kraan-Korteweg05}
Kraan-Korteweg R., 2005, preprint, [astro-ph/0502217]

\bibitem[Land \& Magueijo(2005a)]{LandM05a}
Land K., Magueijo J., 2005, preprint, [astro-ph/0502237]

\bibitem[Land \& Magueijo(2005b)]{LandM05b}
Land K., Magueijo J., 2005, preprint, [astro-ph/0507289]

%\bibitem[{Larson \& Wandelt(2004)}]{Larson2004}
%Larson D.~L., Wandelt B.~D., 2004, \apj, 613, L85

\bibitem[Lynden-Bell et al.(1988)]{Lynden-Bell_88}
Lynden-Bell D. et al., 1988, ApJ, 326, 19

%\bibitem[{Magueijo \& Medeiros(2004)}]{Magueijo_Medeiros}
%Magueijo J., Medeiros J., 2004, \mnras, 351, L1

%\bibitem[{McEwen {et~al.}(2005)McEwen, Hobson, Lasenby, \&
%  Mortlock}]{McEwen2004}
%McEwen J.~D., Hobson M.~P., Lasenby A.~N., Mortlock D.~J., 2005, \mnras, 359,
 % 1583

%\bibitem[{Mukherjee \& Wang(2004)}]{Mukherjee2004}
%Mukherjee P., Wang Y., 2004, \apj, 613, 51

%\bibitem[{{Naselsky} {et~al.}(2005{\natexlab{a}}){Naselsky}, {Chiang},
%  {Olesen}, \& {Novikov}}]{Naselsky2005a}
%{Naselsky} P., {Chiang} L.-Y., {Olesen} P., {Novikov} I., 2005{\natexlab{a}},
 % astro-ph/0505011

%\bibitem[{{Naselsky} {et~al.}(2005{\natexlab{b}}){Naselsky}, {Novikov}, \&
%  {Chiang}}]{Naselsky2005b}
%{Naselsky} P.~D., {Novikov} I.~D., {Chiang} L.-Y., 2005{\natexlab{b}},
%  astro-ph/0506466

%\bibitem[{Park(2004)}]{Park2003}
%Park C.-G., 2004, \mnras, 349, 313

\bibitem[{{Peiris} {et~al.}(2003)}]{Peiris2003}
{Peiris} H.~V., {et~al.}, 2003, \apjs, 148, 213

%\bibitem[{{Prunet} {et~al.}(2005){Prunet}, {Uzan}, {Bernardeau}, \&
%  {Brunier}}]{Prunet}
%{Prunet} S., {Uzan} J., {Bernardeau} F., {Brunier} T., 2005, Phys. Rev., D71,
%  083508
  
%\bibitem[{Schwarz {et~al.}(2004)Schwarz, Starkman, Huterer, \&
%  Copi}]{Schwarz2004}
%Schwarz D.~J., Starkman G.~D., Huterer D., Copi C.~J., 2004, Phys. Rev. Lett., 93,
%  221301

\bibitem[Seljak(1996)]{Seljak96}
Seljak U.,1996 \apj, 463, 1

\bibitem[{{Spergel} {et~al.}(2003)}]{Spergel2003}
{Spergel} D.~N., {et~al.}, 2003, \apjs, 148, 175

\bibitem[Schwarz et al.(2004)]{SchwarzSHC04}
Schwarz D., Starkman G., Huterer D., Copi C., 2004, Phys. Rev. Lett., 93, 22

\bibitem[Sunyaev \& Zeldovich(1972)]{SunyaevZ72}
Sunyaev R., Zeldovich Ya., 1972, 
Comm. Astrophys., Space Phys., 4, 173

\bibitem[Tegmark et al.(2003)]{TegmarkOH03}
Tegmark M., de Oliveira-Costa A., Hamilton A., 2003, Phys. Rev. D68, 123523

%\bibitem[{Tojeiro {et~al.}(2005)Tojeiro, Castro, Heavens, \&
%  Gupta}]{Tojeiro2005}
%Tojeiro R., Castro P.~G., Heavens A.~F., Gupta S., 2005, astro-ph/0507096

\bibitem[Vale \& White(2003)]{ValeW03}
Vale C., White M., 2003, ApJ, 592, 699

%\bibitem[{Vielva {et~al.}(2004)Vielva, Martinez-Gonzalez, Barreiro, Sanz, \&
%  Cayon}]{Vielva2003}
%Vielva P., Martinez-Gonzalez E., Barreiro R.~B., Sanz J.~L., Cayon L., 2004,
%  \apj, 609, 22

%\bibitem[{Wandelt {et~al.}(2004)Wandelt, Larson, \&
%  Lakshminarayanan}]{Wandelt:GEM}
%Wandelt B.~D., Larson D.~L., Lakshminarayanan A., 2004, Phys. Rev., D70, 083511

\bibitem[Zaldarriaga \& Seljak(1998)]{ZaldarriagaS98}
Zaldarriaga M., Seljak U., Phys. Rev. D, 1998, 58, 023003





%\bibitem[Abazajian \& Dodelson(2003)]{AbaDod} 
% Abazajian, K., Dodelson, S., 2003, Phys. Rev. Lett., 91, 041301
%MNRAS, 325, 1065
%Phys. Rep., 340, 291 
%Phys. Rev. D, 64, 3501 
%preprint [astro-ph/0306033]
%A\&A, 338, 375
%AJ, 123, 583 
%ApJ, 600, 17 
%New Astronomy Reviews, 46, 767
%ARA\&A, 37, 127


\end{thebibliography}
\end{document}